\documentclass[twocolumn,pre,superscriptaddress]{revtex4}

\usepackage{dcolumn}
\usepackage{amsmath}

\usepackage{graphicx}

\begin{document}

\renewcommand{\d}{\mathrm{d}}
\newcommand{\Ord}{\mathrm{O}}
\newcommand{\e}{\mathrm{e}}
\newcommand{\half}{\mbox{$\frac12$}}
\newcommand{\set}[1]{\lbrace#1\rbrace}
\newcommand{\av}[1]{\langle#1\rangle}
\newcommand{\etal}{{\it{}et~al.}}
\newcommand{\defn}{\textit}

\newlength{\figurewidth}
\setlength{\figurewidth}{0.95\columnwidth}
\setlength{\parskip}{0pt}
\setlength{\tabcolsep}{6pt}
\setlength{\arraycolsep}{2pt}

\title{Growing distributed networks with arbitrary degree distributions}
\author{Gourab Ghoshal}
\affiliation{Department of Physics, University of Michigan, Ann Arbor, MI
48109}
\affiliation{Michigan Center for Theoretical Physics, University of
Michigan, Ann Arbor, MI, 48109}
\author{M. E. J. Newman}
\affiliation{Department of Physics, University of Michigan, Ann Arbor, MI
48109}
\affiliation{Center for the Study of Complex Systems, University of
Michigan, Ann Arbor, MI 48109}
\begin{abstract}
  We consider distributed networks, such as peer-to-peer networks, whose
  structure can be manipulated by adjusting the rules by which vertices
  enter and leave the network.  We focus in particular on degree
  distributions and show that, with some mild constraints, it is possible
  by a suitable choice of rules to arrange for the network to have any
  degree distribution we desire.  We also describe a mechanism based on
  biased random walks by which appropriate rules could be implemented in
  practice.  As an example application, we describe and simulate the
  construction of a peer-to-peer network optimized to minimize search times
  and bandwidth requirements.
\end{abstract}


\maketitle

\section{Introduction}
\label{intro}
Complex networks, such as the Internet, the worldwide web, and social and
biological networks, have attracted a remarkable amount of attention from
the physics community in recent
years~\cite{Newman03d,Boccaletti06,DM03b,NBW06}.  Most studies of these
systems have concentrated on determining their structure or the effects
that structure has on the behavior of the system.  For instance, a
considerable amount of effort has been devoted to studies of the degree
distributions of networks, their measurement and the formulation of
theories to explain how they come to take the observed forms, and models of
the effect of particular degree distributions on dynamical processes on
networks, network resilience, percolation properties, and many other
phenomena.  Such studies are appropriate for ``naturally occurring''
networks, whose structure grows or is created according to some set of
rules not under our direct control.  The Internet, the web, and social
networks fall into this category even though they are man-made, since their
growth is distributed and not under the control of any single authority.

Not all networks fall in this class however.  There are some networks whose
structure is centrally controlled, such as telephone networks, some
transportation networks, or distribution networks like power grids.  For
these networks it is interesting to ask how, if one can design the network
to have any structure one pleases, one could choose that structure to
optimize some desired property of the network.  For instance,
Paul~\etal~\cite{PTHS04} have considered how the structure of a network
should be chosen to optimize the network's robustness to deletion of its
vertices.

In this paper we study a class of networks that falls between these two
types.  There are some networks that grow in a collaborative, distributed
fashion, so that we cannot control the network's structure directly.  But
we can control some of the rules by which the network forms and this in
turn allows us a limited degree of influence over the structure.  The
archetypal example of such a system is a distributed database such as a
peer-to-peer filesharing network, which is a virtual network of linked
computers that share data among themselves.  The network is formed by a
dynamical process under which individual computers continually join or
leave the network, and the rules of joining and leaving can be manipulated
to some extent by changing the behavior of the software governing
computers' behaviors.  It is well established that the structure of
peer-to-peer networks can have a strong effect on their
performance~\cite{ALPH01,SBR04} but to a large extent that structure has in
the past been regarded as an experimentally determined
quantity~\cite{Hong01}.  Here we consider ways in which the structure can
be manipulated by changing the behavior of individual nodes so as to
optimize network performance.

\section{Growing networks with desired properties} In this paper we focus
primarily on creating networks with desired degree distributions: the
degree distribution typically has a strong effect on the behavior of the
network and is relatively straightforward to treat mathematically.  There
are two basic problems we need to address if we want to create a network
with a specific degree distribution solely by manipulating the rules by
which vertices enter and leave the network.  First, we need to find rules
that will achieve the desired result, and second, we need to find a
practical mechanism that implements those rules and operates in reasonable
time.  We deal with these questions in order.

Our approach to growing a network with a desired degree distribution is
based on the idea of the ``attachment kernel'' introduced by Krapivsky and
Redner~\cite{KR01}.  We assume that vertices join our network at intervals
and that when they do so they form connections---edges---to some number of
other vertices in the network.  By designing the software appropriately, we
can in a peer-to-peer network choose the number of edges a newly joining
vertex makes and also, as we will shortly show, some crucial aspects of
which other vertices those edges connect to.

Let us define $p_k$ to be the degree distribution of our network at some
time, i.e.,~the fraction of vertices having degree~$k$, which satisfies the
normalization condition
\begin{equation}
\sum_{k=0}^\infty p_k = 1.
\end{equation}
And let us define the attachment kernel $\pi_k$ to be the probability that
an edge from a newly appearing vertex connects to a particular preexisting
vertex of degree~$k$, divided by the number~$n$ of vertices in the network.
It is this attachment kernel that we will manipulate to produce a desired
degree distribution.  The extra factor of~$n$ in the definition is not
strictly necessary, but it is convenient: since the total number of
vertices of degree~$k$ is $np_k$, it means that the probability of a new
edge connecting to any vertex of degree~$k$ is just $\pi_kp_k$, and hence
$\pi_k$ satisfies the normalization condition
\begin{equation}
\sum_{k=0}^\infty \pi_k p_k = 1.
\end{equation}

In a peer-to-peer network users may exit the network whenever they want and
we as designers have little control over this aspect of the network
dynamics.  We will assume in the calculations that follow that vertices
simply vanish at random.  We will also assume that, on the typical
time-scales over which people enter and leave the network, the total
size~$n$ of the network does not change substantially, so that the rates at
which vertices enter and leave are roughly equal.  For simplicity let us
say that exactly one vertex enters the network and one leaves per unit time
(although the results presented here are in fact still valid even if only
the probabilities per unit time of addition and deletion of vertices are
equal and not the rates).

Now let us chose the initial degrees of vertices when they join the
network, i.e.,~the number of connections that they form upon entering, at
random from some distribution~$r_k$.  Building on our previous results
in~\cite{MGN06}, we observe that the evolution of the degree distribution
of our network can be described by a rate equation as follows.  The number
of vertices with degree~$k$ at a particular time is~$np_k$.  One unit of
time later we have added one vertex and taken away one vertex, so that the
number with degree~$k$ becomes
\begin{align}
\label{eq:rate}
n p_k' &= np_k + c\pi_{k-1} p_{k-1} - c\pi_k p_k
                \nonumber\\
       &\quad {} + (k+1) p_{k+1} - k p_k - p_k + r_k,
\end{align}
with the convention that $p_{-1}=0$, and $c = \sum_{k=0}^\infty k r_k$,
which is the average degree of vertices added to the network.  The terms $c
\pi_{k-1}p_{k-1}$ and $-c \pi_{k}p_{ k}$ in Eq.~\eqref{eq:rate} represent
the flow of vertices with degree $k-1$ to $k$ and $k$ to $k+1$, as they
gain extra edges with the addition of new vertices.  The terms $(k+1)
p_{k+1}$ and $-k p_k$ represent the flow of vertices with degree $k+1$ and
$k$ to $k$ and $k-1$, as they lose edges with the removal of neighboring
vertices.  The term $-p_k$ represents the probability of removal of a
vertex of degree $k$ and the term $r_k$ represents the addition of a new
vertex with degree $k$ to the network.

Assuming $p_k$ has an asymptotic form in the limit of large time, that form
is given by setting $p_k'=p_k$ thus:
\begin{align}
\label{eq:recur}
c\pi_{k-1} p_{k-1} - c\pi_k p_k + (k+1) p_{k+1} &- k p_k \nonumber\\
    & {} - p_k + r_k = 0.
\end{align}
Following previous convention~\cite{NSW01}, let us define a generating
function $G_0(z)$ for the degree distribution thus:
\begin{equation}
G_0(z) =  \sum_{k=0}^\infty p_k z^k,
\end{equation}
as well as generating functions for the degrees of vertices added and for
the attachment kernel thus:
\begin{eqnarray}
F(z) &=& \sum_{k=0}^\infty r_k z^k, \\
H(z) &=& \sum_{k=0}^\infty \pi_k p_k z^k.
\end{eqnarray}
Multiplying both sides of~\eqref{eq:recur} by $z^k$ and summing over~$k$,
we then find that the generating functions satisfy the differential
equation
\begin{equation}
\label{eq:diffeq}
(1-z) {\d G_0\over\d z} - G_0(z) - c(1-z) H(z) + F(z) = 0.
\end{equation}
We are interested in creating a network with a given degree distribution,
i.e.,~with a given $G_0(z)$.  Rearranging~\eqref{eq:diffeq}, we find that
the choice of attachment kernel~$\pi_k$ that achieves this is such that
\begin{equation}
\label{eq:condition1}
H(z) = {1\over c} \biggl[ {\d G_0\over\d z} + {F(z)-G_0(z)\over1-z} \biggr].
\end{equation}
Taking the limit $z\to1$, noting that normalization requires that all the
generating functions tend to~1 at $z=1$, and applying L'Hopital's rule, we
find
\begin{equation}
1 = \frac{1}{c}\left[\av{k} + \av{k} - c\right],
\end{equation}
where we have made use of the fact that the average degree in the network
is $\av{k}=G_0'(1)$ and $c=H'(1)$.  Rearranging, we then find that
$c=\av{k}$.  In other words, solutions to Eq.~\eqref{eq:diffeq} require
that the average degree~$c$ of vertices added to the network be equal to
the average degree of vertices in the network as a whole.  Making use of
this result, we can write Eq.~\eqref{eq:condition1} in the form
\begin{equation}
\label{eq:condition2}
H(z) = G_1(z) + {F(z)-G_0(z)\over c(1-z)},
\end{equation}
where $G_1(z) = G_0'(z)/G_0'(1) = \sum_k q_k z^k$ is the generating function
for the so-called excess degree distribution
\begin{equation}
\label{eq:excess}
q_k = {(k+1)p_{k+1}\over\av{k}},
\end{equation}
which appears in many other network-related calculations---see, for
instance, Ref.~\cite{NSW01}.

Now it is straightforward to derive the desired attachment kernel.  Noting
that
\begin{equation}
{1\over1-z} = \sum_{k=0}^\infty z^k,
\end{equation}
we can simply read off the coefficient of $z^k$ on either side of
Eq.~\eqref{eq:condition2}, to give
\begin{equation}
\pi_k p_k = q_k + {1\over c} \sum_{m=0}^k (r_m-p_m),
\end{equation}
or equivalently
\begin{equation}
\label{eq:genpik}
\pi_k = {1\over cp_k} \bigl[ (k+1) p_{k+1} + P_{k+1} - R_{k+1} \bigr],
\end{equation}
where $P_k$ is the cumulative distribution of vertex degrees and $R_k$ is
the cumulative distribution of added degrees:
\begin{equation}
P_k = \sum_{m=k}^\infty p_m,\qquad R_k = \sum_{m=k}^\infty r_m.
\end{equation}

Since we are at liberty to choose both $r_k$ and~$\pi_k$, we have many
options for satisfying Eq.~\eqref{eq:genpik}; given (almost) any choice of
the distribution~$r_k$ of the degrees of added vertices, we can find the
corresponding $\pi_k$ that will give the desired final degree distribution
of the network.  One simple choice would be to make the degree distribution
of the added vertices the same as the desired degree distribution, so that
$R_k=P_k$.  Then
\begin{equation}
\label{eq:pik}
\pi_k = {q_k\over p_k} = {(k+1)p_{k+1}\over cp_k}.
\end{equation}
In other words, if we have some desired degree distribution~$p_k$ for our
network, one way to achieve it is to add vertices with exactly that degree
distribution and then arrange the attachment process so that the degree
distribution remains preserved thereafter, even as vertices and edges are
added to and removed from the network.  Equation~\eqref{eq:pik} tells us
the choice of attachment kernel that will achieve this.
Equation~\eqref{eq:pik} will work for essentially any choice of degree
distribution~$p_k$, except choices for which $p_k=0$ and $p_{k+1}>0$ for
some~$k$.  In the latter case Eq.~\eqref{eq:pik} will diverge for some
value(s) of~$k$.

\subsection{Example: power-law degree distribution}
As an example, consider the creation of a network with a power-law degree
distribution.  Adamic~\etal~\cite{ALPH01} have shown that search processes
on peer-to-peer networks with power-law degree distributions are
particularly efficient, so there are reasons why one might want to generate
such a network.

Let us choose
\begin{equation}
p_k = \begin{cases}
      \enspace C k^{-\gamma} & \text{for $k\ge1$,} \\
      \enspace p_0           & \text{for $k=0$,}
      \end{cases}
\end{equation}
where $\gamma$ and $p_0$ are constants and the normalizing factor~$C$ is
given by
\begin{equation}
C = {1-p_0\over\zeta(\gamma)},
\end{equation}
where $\zeta(\gamma)$ is the Riemann zeta-function.  Then the mean degree
is
\begin{equation}
\av{k} = c = (1-p_0) {\zeta(\gamma-1)\over\zeta(\gamma)},
\end{equation}
and Eq.~\eqref{eq:pik} tells us that the correct choice of attachment
kernel in this case is
\begin{equation}
\label{eq:plpik}
\pi_k = {1\over1-p_0}\,{\zeta(\gamma)\over\zeta(\gamma-1)}\,
        {k^\gamma\over(k+1)^{\gamma-1}},
\end{equation}
for $k\ge1$ and
\begin{equation}
\pi_0 = {1\over p_0\zeta(\gamma-1)}.
\end{equation}

It is interesting to note that as $k$ becomes large, this attachment kernel
goes as $\pi_k\sim k$, the so-called (linear) preferential attachment form
in which vertices connect to others in simple proportion to their current
degree.  In growing networks this form is known to give rise,
asymptotically, to a power-law degree distribution.  It is important to
understand, however, that in the present case the network is not growing
and hence, despite the apparent similarity, this is not the same result.
Indeed, it is known that for non-growing networks, purely linear
preferential attachment does not produce power-law degree
distributions~\cite{SR04b,CFV04}, but instead generates stretched
exponential distributions~\cite{MGN06}.  Thus it is somewhat surprising to
observe that one can, nonetheless, create a power-law degree distribution
in a non-growing network using an attachment kernel that seems,
superficially, quite close to the linear form.

Sarshar and Roychowdhury~\cite{SR04b} showed previously that it is possible
to generate a non-growing power-law network by using linear preferential
attachment and then compensating for the expected loss of power-law
behavior by rewiring the connections of some vertices after their addition
to the network.  Our results indicate that, although this process will
certainly work, it is not necessary: a slight modification to the
preferential attachment process will achieve the same goal and frees us
from the need to rewire any edges.

Note also that~\eqref{eq:plpik} is not the only solution of
Eq.~\eqref{eq:genpik} that will generate a power-law distribution.  If we
choose a different (e.g.,~non-power-law) distribution for the vertices
added to the network, we can still generate an overall power-law
distribution by choosing the attachment kernel to satisfy
Eq.~\eqref{eq:genpik}.  Suppose, for instance, that, rather than adding
vertices with a power-law degree distribution, we prefer to give them a
Poisson distribution with mean~$c$:
\begin{equation}
r_k = \e^{-c}\,{c^k\over k!}.
\end{equation}
In this case $R_k = 1 - \Gamma(k,c)/\Gamma(k)$, where $\Gamma(k)$ is the
standard gamma function and $\Gamma(k,c)$ is the incomplete gamma function.
Then the power law is correctly generated by the choice
\begin{align}
\pi_k &= {1\over1-p_0}\,{\zeta(\gamma)\over\zeta(\gamma-1)}\,k^\gamma
         \biggl[ (k+1)^{-\gamma+1} + \zeta(\gamma,k+1) \nonumber\\
      &\qquad - {\zeta(\gamma)\over1-p_0}
            \biggl( 1 - {\Gamma(k+1,c)\over\Gamma(k+1)} \biggr) \biggr],
\end{align}
for $k\ge1$, where $\zeta(\gamma,x)$ is the generalized zeta function
$\zeta(\gamma,x) = \sum_{k=0}^\infty (k+x)^{-\gamma}$.  For $k=0$,
\begin{equation}
\pi_0 = {1\over p_0\zeta(\gamma-1)}
        \biggl[ 1 + {\e^{-c}-p_0\over1-p_0}\,\zeta(\gamma) \biggr].
\end{equation}

\section{A practical implementation}
\label{sec:walks}
In theory, we should be able use the ideas of the previous section to grow
a network with a desired degree distribution.  This does not, however, yet
mean we can do so in practice.  To make our scheme a practical reality, we
still need to devise a realistic way to place edges between vertices with
the desired attachment kernel~$\pi_k$.  If each vertex entering the network
knew the identities and degrees of all other vertices, this would be easy:
we would simply select a degree~$k$ at random in proportion to~$\pi_k p_k$,
and then attach our new edge to a vertex chosen uniformly at random from
those having that degree.

In the real world, however, and particularly in peer-to-peer networks, no
vertex ``knows'' the identity of all others.  Typically, computers only
know the identities (such as IP addresses) of their immediate network
neighbors.  To get around this problem, we propose the following scheme,
which makes use of biased random walks.

A random walk, in this context, is a succession of steps along edges in our
network where at each vertex~$i$ we choose to step next to a vertex chosen
at random from the set of neighbors of~$i$.  In the context of a
peer-to-peer computer network, for example, such a walk can be implemented
by message passing between peers.  The ``walker'' is a message or data
packet that is passed from computer to neighboring computer, with each
computer making random choices about which neighbor to pass to next.

Starting a walk from any vertex in the network, we can sample vertices by
allowing the walk to take some fixed number of steps and then choosing the
vertex that it lands upon on its final step.  We will consider random walks
in which the choice of which step to make at each vertex is deliberately
biased to create a desired probability distribution for the sample as
follows.

Consider a walk in which a walker at vertex~$j$ chooses uniformly at random
one of the $k_j$ neighbors of that vertex.  Let us call this neighbor~$i$.
Then the walk takes a step to vertex~$i$ with some acceptance probability
$P_{ij}$.  The total probability~$T_{ij}$ of a transition from $j$ to $i$
given that we are currently at~$j$ is
\begin{equation}
\label{eq:tij}
T_{ij} = {A_{ij}\over k_j} P_{ij},
\end{equation}
where $k_j$ is the degree of vertex~$j$ and $A_{ij}$ is an element of the
adjacency matrix:
\begin{equation}
A_{ij} = \begin{cases}
           \enspace 1 & \text{if there is an edge joining vertices $i,j$,} \\
           \enspace 0 & \text{otherwise.}
         \end{cases}
\label{eq:adjacency}
\end{equation}
If the step is not accepted, then the random walker remains at vertex~$j$
for the current step.

This random walk constitutes an ordinary Markov process, which converges to
a distribution~$p_i$ over vertices provided the network is connected
(i.e.,~consists of a single component) and provided $T_{ij}$ satisfies the
detailed balance condition
\begin{equation}
T_{ij} p_j = T_{ji} p_i.
\end{equation}

In the present case we wish to select vertices in proportion to the
attachment kernel~$\pi_k$.  Setting $p_i=\pi_{k_i}$, this implies that
$T_{ij}$ should satisfy
\begin{equation}
{T_{ij}\over T_{ji}} = {p_i\over p_j} = {\pi_{k_i}\over\pi_{k_j}}.
\end{equation}
Or, making use of Eqs.~\eqref{eq:pik} and~\eqref{eq:tij} for the case where
$r_k=p_k$, we find
\begin{equation}
{P_{ij}\over P_{ji}} = {(k_i+1) p_{k_i+1}\over k_i p_{k_i}} \,
                       {k_j p_{k_j}\over(k_j+1) p_{k_j+1}}
                     = {q_{k_i}q_{k_j-1}\over q_{k_j}q_{k_i-1}},
\end{equation}
where $q_k$ is again the excess degree distribution, Eq.~\eqref{eq:excess}.

In practice, we can satisfy this equation by making the standard
Metropolis-Hastings choice for the acceptance probability:
\begin{equation}
P_{ij} = \begin{cases}
         \enspace q_{k_i}q_{k_j-1}/q_{k_j}q_{k_i-1} &
         \text{if $q_{k_i}/q_{k_i-1}<q_{k_j}/q_{k_j-1}$,} \\
         \enspace 1      & \text{otherwise.}
         \end{cases}
\end{equation}
Thus the calculation of the acceptance probability requires only that each
vertex know the degrees of its neighboring vertices, which can be
established by a brief exchange of data when the need arises.

As an example, suppose we wish to generate a network with a Poisson degree
distribution
\begin{equation}
\label{eq:poissonmu}
p_k = \e^{-\mu}\,{\mu^k\over k!},
\end{equation}
where $\mu$ is the mean of the Poisson distribution.  Then we find that the
appropriate choice of acceptance ratio is
\begin{equation}
\label{eq:poissonrule}
P_{ij} = \begin{cases}
         \enspace k_j/k_i & \text{if $k_i>k_j$,} \\
         \enspace 1       & \text{otherwise.}
         \end{cases}
\end{equation}
(As discussed above, we must also make sure to choose the mean degree~$c$
of vertices added to the network to be equal to~$\mu$.)

Our proposed method for creating a network is thus as follows.  Each newly
joining vertex~$i$ first chooses a degree~$k$ for itself, which is drawn
from the desired distribution~$p_k$.  It must also locate one single other
vertex~$j$ in the network.  It might do this for instance using a list of
known previous members of the network or a standardized list of permanent
network members.  Vertex~$j$ is probably not selected randomly from the
network, so it is \emph{not} chosen as a neighbor of~$i$.  Instead, we use
it as the starting point for a set of~$k$ biased random walkers of the type
described above.  Each walker consists of a message, which starts at~$j$
and propagates through the network by being passed from computer to
neighboring computer.  The message contains (at a minimum) the address of
the computer at vertex~$i$ as well as a counter that is updated by each
computer to record the number of steps the walker has taken.  (Bear in mind
that steps on which the walker doesn't move, because the proposed move was
rejected, are still counted as steps.)  The computer that the walker
reaches on its $t$th step, where $t$ is a fixed but generous constant
chosen to allow enough time for mixing of the walk, establishes a new
network edge between itself and vertex~$i$ and the walker is then deleted.
When all $k$ walkers have terminated in this way, vertex~$i$ has $k$ new
neighbors in the network, chosen in proportion to the correct attachment
kernel~$\pi_k$ for the desired distribution.  After a suitable interval of
time, this process will result in a network that has the chosen degree
distribution~$p_k$, but is otherwise random.

\begin{figure}[t]
\includegraphics[width=\figurewidth]{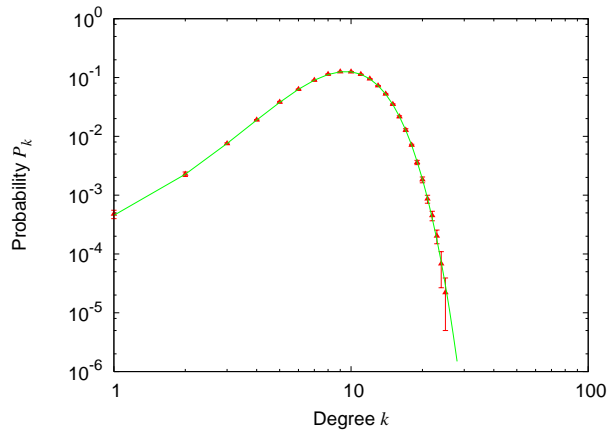}
\caption{The degree distribution for a network of $n=50\,000$ vertices
  generated using the biased random walk mechanism described in the text
  with $\mu=10$.  The points represent the results of our simulations and
  the solid line is the target distribution, Eq.~\eqref{eq:poissonmu}.}
\label{fig:degdist}
\end{figure}

As a test of this method, we have performed simulations of the growth of a
network with a Poisson degree distribution as in
Eq.~\eqref{eq:poissonrule}.  Starting from a random graph of the desired
size~$n$, we randomly add and remove vertices according to the prescription
given above.  Figure~\ref{fig:degdist} shows the resulting degree
distribution for the case $\mu=10$, along with the expected Poisson
distribution.  As the figure shows, the agreement between the two is
excellent.

\section{Example application}

As an example of the application of these ideas we consider peer-to-peer
networks.  Bandwidth restrictions and search times place substantial
constraints on the performance of peer-to-peer networks, and the methods of
the previous sections can be used to nudge networks towards a structure
that improves their performance in these respects.  More sophisticated
applications are certainly possible, but the one presented here offers an
indication of the kinds of possibilities open to us.

\subsection{Definition of the problem}
Consider a distributed database consisting of a set of computers each of
which holds some data items.  Copies of the same item can exist on more
than one computer, which would make searching easier, but we will not
assume this to be the case.  Computers are connected together in a
``virtual network,'' meaning that each computer is designated as a
``neighbor'' of some number of other computers.  These connections between
computers are purely notional: every computer can communicate with every
other directly over the Internet or other physical network.  The virtual
network is used only to limit the amount of information that computers have
to keep about their peers.

Each computer maintains a directory of the data items held by its network
neighbors, but not by any other computers in the network.  Searches for
items are performed by passing a request for a particular item from
computer to computer until it reaches one in whose directory that item
appears, meaning that one of that computer's neighbors holds the item.  The
identity of the computer holding the item is then transmitted back to the
origin of the search and the origin and target computers communicate
directly thereafter to negotiate the transfer of the item.  This basic
model is essentially the same as that used by other authors~\cite{ALPH01}
as well as by many actual peer-to-peer networks in the real world.  Note
that it achieves efficiency by the use of relatively large directories at
each vertex of the network, which inevitably use up memory resources on the
computers.  However, with standard hash-coding techniques and for databases
of the typical sizes encountered in practical situations (thousands or
millions of items) the amounts of memory involved are quite modest by
modern standards.

\subsection{Search time and bandwidth}
The two metrics of search performance that we consider in this example are
\defn{search time} and \defn{bandwidth}, both of which should be low for a
good search algorithm.  We define the search time to be the number of steps
taken by a propagating search query before the desired target item is
found.  We define the bandwidth for a vertex as the average number of
queries that pass through that vertex per unit time.  Bandwidth is a
measure of the actual communications bandwidth that vertices must expend to
keep the network as a whole running smoothly, but it is also a rough
measure of the CPU time they must devote to searches.  Since these are
limited resources it is crucial that we not allow the bandwidth to grow too
quickly as vertices are added to the network, otherwise the size of the
network will be constrained, a severe disadvantage for networks that can in
some cases swell to encompass a significant fraction of all the computers
on the planet.  (In some early peer-to-peer networks, issues such as this
did indeed place impractical limits on network size~\cite{Ritter00,RFI02}.)

Assuming that the average behavior of a user of the database remains
essentially the same as the network gets larger, the number of queries
launched per unit time should increase linearly with the size of the
network, which in turn suggests that the bandwidth per vertex might also
increase with network size, which would be a bad thing.  As we will show,
however, it is possible to avoid this by designing the topology of the
network appropriately.

\subsection{Search strategies and search time}
\label{strategy}
In order to treat the search problem quantitatively, we need to define a
search strategy or algorithm.  Here we consider a very simple---even
brainless---strategy, again employing the idea of a random walk.  This
\defn{random walk search} is certainly not the most efficient strategy
possible, but it has two significant advantages for our purposes.  First,
it is simple enough to allow us to carry out analytic calculations of its
performance.  Second, as we will show, even this basic strategy can be made
to work very well.  Our results constitute an existence proof that good
performance is achievable: searches are necessarily possible that are at
least as good as those analyzed here.

The definition of our random walk search is simple: the vertex~$i$
originating a search sends a query for the item it wishes to find to one of
its neighbors~$j$, chosen uniformly at random.  If that item exists in the
neighbor's directory the identity of the computer holding the item is
transmitted to the originating vertex and the search ends.  If not, then
$j$ passes the query to one of its neighbors chosen at random, and so
forth.  (One obvious improvement to the algorithm already suggests itself:
that $j$ not pass the query back to $i$ again.  As we have said, however,
our goal is simplicity and we will allow such ``backtracking'' in the
interests of simplifying the analysis.)

We can study the behavior of this random walk search by a method similar to
the one we employed for the analysis of the biased random walks of
Section~\ref{sec:walks}.  Let $p_i$ be the probability that our random
walker is at vertex~$i$ at a particular time.  Then the probability~$p_i'$
of its being at $i$ one step later, assuming the target item has not been
found, is
\begin{equation}
p_i' = \sum_j {A_{ij}\over k_j} p_j,
\label{eq:rw1}
\end{equation}
where $k_j$ is the degree of vertex~$j$ and $A_{ij}$ is an element of the
adjacency matrix, Eq.~\eqref{eq:adjacency}.  Under the same conditions as
before the probability distribution over vertices then tends to the fixed
point of~\eqref{eq:rw1}, which is at
\begin{equation}
p_i = {k_i\over2m},
\label{eq:pi}
\end{equation}
where $m$ is the total number of edges in the network.  That is, the random
walk visits vertices with probability proportional to their degrees.  (An
alternative statement of the same result is that the random walk visits
\emph{edges} uniformly.)


When our random walker arrives at a previously unvisited vertex of
degree~$k_i$, it ``learns'' from that vertex's directory about the items
held by all immediate neighbors of the vertex, of which there are~$k_i-1$
excluding the vertex we arrived from (whose items by definition we already
know about).  Thus at every step the walker gathers more information about
the network.  The average number of vertices it learns about upon making a
single step is $\sum_i p_i(k_i-1)$, with $p_i$ given by~\eqref{eq:pi}, and
hence the total number it learns about after~$\tau$ steps is
\begin{equation}
{\tau\over2m} \sum_i k_i(k_i-1) = \tau \biggl[ {\av{k^2}\over\av{k}} - 1
                                       \biggr],
\end{equation}
where $\av{k}$ and $\av{k^2}$ represent the mean and mean-square degrees in
the network and we have made use of $2m=n\av{k}$.  (There is in theory a
correction to this result because the random walker is allowed to backtrack
and visit vertices visited previously.  For a well-mixed walk, however,
this correction is of order $1/\av{k}$, which, as we will see, is
negligible for the networks we will be considering.)

How long will it take our walker to find the desired target item?  That
depends on how many instances of the target exist in the network.  In many
cases of practical interest, copies of items exist on a fixed
\emph{fraction} of the vertices in the network, which makes for quite an
easy search.  We will not however assume this to be the case here.  Instead
we will consider the much harder problem in which copies of the target item
exist on only a fixed \emph{number} of vertices, where that number could
potentially be just~1.  In this case, the walker will need to learn about
the contents of $\Ord(n)$ vertices in order to find the target and hence
the average time to find the target is given by
\begin{equation}
\tau \biggl[ {\av{k^2}\over\av{k}} - 1 \biggr] = An,
\end{equation}
for some constant~$A$, or equivalently,
\begin{equation}
\tau = A {n\over\av{k^2}/\av{k} - 1}.
\label{eq:tau}
\end{equation}

Consider, for instance, a network with a power-law degree distribution of
the form, $p_k = C k^{-\gamma}$, where $\gamma$ is a positive exponent and
$C$ is a normalizing constant chosen such that $\sum_{k=0}^\infty p_k = 1$.
Real-world networks usually exhibit power-law behavior only over a certain
range of degree.  Taking the minimum of this range to be $k=1$ and denoting
the maximum by $k_\mathrm{max}$, we have
\begin{equation}
{\av{k^2}\over\av{k}} \sim
  - \frac{k_\mathrm{max}^{3-\gamma}-1}{k_\mathrm{max}^{2-\gamma}-1}.
\end{equation}
Typical values of the exponent~$\gamma$ fall in the range $2<\gamma<3$, so
that $k_\mathrm{max}^{2-\gamma}$ is small for large $k_\mathrm{max}$ and
can be ignored.  On the other hand, $k_\mathrm{max}^{3-\gamma}$~becomes
large in the same limit and hence $\av{k^2}/\av{k}\sim
k_\mathrm{max}^{3-\gamma}$ and
\begin{equation}
\tau \sim n k_\mathrm{max}^{\gamma-3}.
\end{equation}
The scaling of the search time with system size~$n$ thus depends, in this
case, on the scaling of the maximum degree~$k_\mathrm{max}$.

As an example, Aiello~\etal~\cite{ACL00} studied power-law degree
distributions with a cut-off of the form $k_\mathrm{max} \sim
n^{1/\gamma}$, which gives
\begin{equation}
\tau \sim n^{2-3/\gamma}.
\end{equation}
A similar result was obtained previously by Adamic~\etal~\cite{ALPH01}
using different methods.

\subsection{Bandwidth}
Bandwidth is the mean number of queries reaching a given vertex per unit
time.  Equation~\eqref{eq:pi} tells us that the probability that a
particular current query reaches vertex~$i$ at a particular time is
$k_i/2m$, and assuming as discussed above that the number of queries
initiated per unit time is proportional to the total number of vertices,
the bandwidth for vertex~$i$ is
\begin{equation}
Bn{k_i\over2m} = B{k_i\over\av{k}},
\end{equation}
where $B$ is another constant.

This implies that high-degree vertices will be overloaded by comparison
with low-degree ones so that, despite their good performance in terms of
search times, networks with power-law or other highly right-skewed degree
distributions may be undesirable in terms of bandwidth, with bottlenecks
forming around the vertices of highest degree that could harm the
performance of the entire network.  If we wish to distribute load more
evenly among the computers in our network, a network with a tightly peaked
degree distribution is desirable.

\subsection{Choice of network}
\label{model}
A simple and attractive choice for our network is the Poisson distributed
network of Section~\ref{sec:walks}.  For a Poisson degree distribution with
mean~$\mu$ we have $\av{k}=\mu$ and $\av{k^2}=\mu^2+\mu$.  Then, using
Eq.~\eqref{eq:tau}, the average search time is
\begin{equation}
\tau = A {n\over\mu}.
\label{eq:tau1}
\end{equation}
As we have seen, a network of this type can be realized in practice with a
biased-random-walker attachment mechanism of the kind described in
Section~\ref{sec:walks}.

Now if we allow $\mu$ to grow as some power of the size of the entire
network, $\mu\sim n^\alpha$ with $0\leq\alpha\leq1$, then $\tau \sim
n^{1-\alpha}$.  For smaller values of~$\alpha$, searches will take longer,
but vertices' degrees are lower on average meaning that each vertex will
have to devote less memory resources to maintaining its directory.
Conversely, for larger~$\alpha$, searches will be completed more quickly at
the expense of greater memory usage.  In the limiting case $\alpha=1$,
searches are completed in constant time, independent of the network size,
despite the simple-minded nature of the random walk search algorithm.

The price we pay for this good performance is that the network becomes
dense, having a number of edges scaling as~$n^{1+\alpha}$.  It is important
to bear in mind, however, that this is a \emph{virtual} network, in which
the edges are a purely notional construct whose creation and maintenance
carries essentially zero cost.  There \emph{is} a cost associated with the
directories maintained by vertices, which for $\alpha=1$ will contain
information on the items held by a fixed fraction of all the vertices in
the network.  For instance, each vertex might be required to maintain a
directory of 1\% of all items in the network.  Because of the nature of
modern computer technology, however, we don't expect this to create a
significant problem.  User time (for performing searches) and CPU time and
bandwidth are scarce resources that must be carefully conserved, but memory
space on hard disks is cheap, and the tens or even hundreds of megabytes
needed to maintain a directory is considered in most cases to be a small
investment.  By making the choice $\alpha=1$ we can trade cheap memory
resources for essentially optimal behavior in terms of search time and this
is normally a good deal for the user.

We note also that the search process is naturally parallelizable: there is
nothing to stop the vertex originating a search from sending out several
independent random walkers and the expected time to complete the search
will be reduced by a factor of the number of walkers.  Alternatively, we
could reduce the degrees of all vertices in the network by a constant
factor and increase the number of walkers by the same factor, which would
keep the average search time constant while reducing the sizes of the
directories substantially, at the cost of increasing the average bandwidth
load on each vertex.

\begin{figure}[t]
\includegraphics[width=\figurewidth]{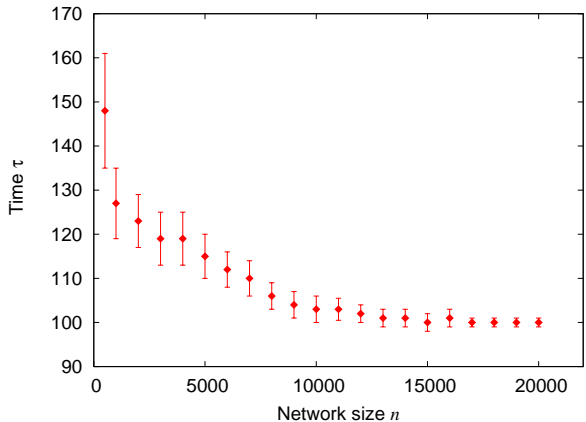}
\caption{The time $\tau$ for the random walk search to find an item
deposited at a random vertex, as a function of the number of vertices~$n$.}
\label{fig:search}
\end{figure}

As a test of our proposed search scheme, we have performed simulations of
the procedure on Poisson networks generated using the random-walker method
of Section~\ref{sec:walks}.  Figure~\ref{fig:search} shows as a function of
network size the average time~$\tau$ taken by a random walker to find an
item placed at a single randomly chosen vertex in the network.  As we can
see, the value of~$\tau$ does indeed tend to a constant (about 100 steps in
this case) as network size becomes large.

We should also point out that for small values of $\mu$ vertices with
degree zero could cause a problem.  A vertex that loses all of its edges
because its neighbors have all left the network can no longer be reached by
our random walkers, and hence no vertices can attach to them and our
attachment scheme breaks down.  However, in the case considered here, where
$\mu$ becomes large, the number of such vertices is exponentially small,
and hence they can be neglected without substantial deleterious effects.
Any vertex that does find itself entirely disconnected from the network can
simply rejoin by the standard mechanism.

\subsection{Item frequency distribution} In most cases, the search problem
posed above is not a realistic representation of typical search problems
encountered in peer-to-peer networks.  In real networks, copies of items
often occur in many places in the network.  Let $s$ be the number of times
a particular item occurs in the network and let $p_s$ be the probability
distribution of $s$ over the network, i.e.,~$p_s$ is the fraction of items
that exist in~$s$ copies.

If the item we are searching for exists in~$s$ copies, then
Eq.~\eqref{eq:tau1} becomes
\begin{equation}
\tau_s = A {n\over\mu s},
\end{equation}
since the chance of finding a copy of the desired item is multiplied by~$s$
on each step of the random walk.  On the other hand, it is likely that the
frequency of searches for items is not uniformly distributed: more popular
items, that is those with higher~$s$, are likely to be searched for more
often than less popular ones.  For the purposes of illustration, let us
make the simple assumption that the frequency of searches for a particular
item is proportional to the item's popularity.  Then the average time taken
by a search is
\begin{equation}
\av{\tau} = {\sum_{s=1}^\infty sp_s\tau_s\over\sum_{s=1}^\infty sp_s}
          = A {n\over \mu\av{s}},
\label{eq:tau2}
\end{equation}
where we have made use of $\sum_s p_s = 1$ and $\sum_s sp_s = \av{s}$.

One possibility is that the total number of copies of items in the network
increases in proportion to the number of vertices, but that the number of
\emph{distinct} items remains roughly the same, so that the average number
of copies of a particular item increases as $\av{s}\sim n$.  In this case,
$\av{\tau}$~becomes independent of~$n$ even when $\mu$ is constant, since
we have to search only a constant number of vertices, not a constant
fraction, to find a desired item.  Perhaps a more realistic possibility is
that the number of distinct items increases with network size, but does so
slower than~$n$, in which case one can achieve constant search times with a
mean degree~$\mu$ that also increases slower than~$n$, so that directory
sizes measured as a fraction of the network size dwindle.

An alternative scenario is one of items with a power-law frequency
distribution~$p_s\sim s^{-\delta}$.  This case describes, for example, most
forms of mass art or culture including books and recordings, emails and other
messages circulating on the Internet, and many others~\cite{Newman05b}.
The mean time to perform a search in the network then depends on the value
of the exponent~$\delta$.  In many cases we have $\delta>2$, which means
that $\av{s}$ is finite and well-behaved as the database becomes large, and
hence~$\av{\tau}$, Eq.~\eqref{eq:tau2}, differs from Eq.~\eqref{eq:tau1} by
only a constant factor.  (That factor may be quite large, making a
significant practical difference to the waiting time for searches to
complete, but the scaling with system size is unchanged.)  If $\delta<2$,
however, then $\av{s}$ becomes ill-defined, having a formally divergent
value, so that $\av{\tau}\to0$ as system size becomes large.  Physically,
this represents the case in which most searches are for the most commonly
occurring items, and those items occur so commonly that most searches
terminate very quickly.

While this extra speed is a desirable feature of the search process, it's
worth noting that average search time may not be the most important metric
of performance for users of the network.  In many situations, worst-case
search time is a better measure of the ability of the search algorithm to
meet users' demands.  Assuming that the most infrequently occurring items
in the network occur only once, or only a fixed number of times, the
worst-case performance will still be given by Eq.~\eqref{eq:tau1}.

\subsection{Estimating network size} One further detail remains to be
considered.  If we want to make the mean degree~$\mu$ of vertices added to
the network proportional to the size~$n$ of the entire network, or to some
power of~$n$, we need to know~$n$, which presents a challenge since, as we
have said, we do not expect any vertex to know the identity of all or even
most of the other the vertices.  This problem can be solved using a
breadth-first search, which can be implemented once again by message
passing across the network.  One vertex~$i$ chosen at random (or more
realistically every vertex, at random but stochastically constant intervals
proportional to system size) sends messages to some number~$d$ of randomly
chosen neighbors.  The message contains the address of vertex~$i$, a unique
identifier string, and a counter whose initial value is zero.  Each
receiving vertex increases the counter by~1, passes the message on to one
of its neighbors, and also sends messages with the same address,
identifier, and with counter zero to $d-1$ other neighbors.  Any vertex
receiving a message with an identifier it has seen previously sends the
value of the counter contained in that message back to vertex~$i$, but does
not forward the message to any further vertices.  If vertex~$i$ adds
together all of the counter values it receives, the total will equal the
number of vertices (other than itself) in the entire network.  This number
can then be broadcast to every other vertex in the network using a similar
breadth-first search (or perhaps as a part of the next such search
instigated by vertex~$i$.)

The advantage of this process is that it has a total bandwidth cost (total
number of messages sent) equal to~$dn$.  For constant~$d$ therefore, the
cost per vertex is a constant and hence the process will scale to
arbitrarily large networks without consuming bandwidth.  The (worst-case)
time taken by the process depends on the longest geodesic path between any
two vertices in the network, which is $\Ord(\log n)$.  Although not as good
as $\Ord(1)$, this still allows the network to scale to exponentially large
sizes before the time needed to measure network size becomes an issue, and
it seems likely that directory size (which scales linearly with or as a
power of~$n$ depending on the precise algorithm) will become a limiting
factor long before this happens.

\section{Conclusions}
In this paper, we have considered the problem of designing networks
indirectly by manipulating the rules by which they evolve.  For certain
types of networks, such as peer-to-peer networks, the limited control that
this manipulation gives us over network structure, such as the ability to
impose an arbitrary degree distribution of our choosing on the network, may
be sufficient to generate significant improvements in network performance.
Using generating function methods, we have shown that it is possible to
impose a (nearly) arbitrary degree distribution on a network by appropriate
choice of the ``attachment kernel'' that governs how newly added vertices
connect to the network.  Furthermore, we have described a scheme based on
biased random walks whereby arbitrary attachment kernels can be implemented
in practice.

We have also considered what particular choices of degree distribution
offer the best performance in idealized networks under simple assumptions
about search strategies and bandwidth constraints.  We have given general
formulas for search times and bandwidth usage per vertex and studied in
detail one particularly simple case of a Poisson network that can be
realized in straightforward fashion using our biased random walker scheme,
allows us to perform decentralized searches in constant time, and makes
only constant bandwidth demands per vertex, even in the limit where the
database becomes arbitrarily large.  No part of the scheme requires any
centralized knowledge of the network, making the network a true
peer-to-peer network, in the sense of having client nodes only and no
servers.

One important issue that we have neglected in our discussion is that of
``supernodes'' in the network.  Because the speed of previous search
strategies has been recognized as a serious problem for peer-to-peer
networks, designers of some networks have chosen to designate a subset of
network vertices (typically those with above-average bandwidth and CPU
resources) as supernodes.  These supernodes are themselves connected
together into a network over which all search activity takes place.  Other
client vertices then query this network when they want a search performed.
Since the size of the supernode network is considerably less than the size
of the network as a whole, this tactic increases the speed of searches,
albeit only by a constant factor, at the expense of heavier load on the
supernode machines.  It would be elementary to generalize our approach to
incorporate supernodes.  One would simply give each supernode a directory
of the data items stored by the client vertices of its supernode neighbors.
Then searches would take place exactly as before, but on the supernode
network alone, and client vertices would query the supernode network to
perform searches.  In all other respects the mechanisms would remain the
same.

\begin{acknowledgments}
The authors thank Lada Adamic for useful discussions.  This work was funded
in part by the National Science Foundation under grant number DMS--0405348
and by the James S. McDonnell Foundation.
\end{acknowledgments}

\end{document}